\documentclass[prd,twocolumn,nofootinbib,showpacs]{revtex4}%
\usepackage{epsfig,graphicx,bm}
\usepackage[latin2]{inputenc}
\usepackage{color}
\usepackage{t1enc}
\usepackage{amssymb}
\usepackage{amsmath}
\usepackage{graphicx}
\usepackage{bm}
\usepackage{amsfonts}%
\begin{document}
\title{Could the OPERA setup send a bit of information faster than
  light?}  \author{$\text{F.~Giacosa}^{1}$}
\author{$\text{P.~Kov\'acs}^{1,2}$} \author{$\text{S.~Lottini}^{1}$}
\affiliation{$^{1}$Institute for Theoretical Physics, Johann Wolfgang
  Goethe University, Max-von-Laue-Str.~1, D-60438 Frankfurt am Main}
\affiliation{$^{2}$Research Institute for Particle and Nuclear Physics
  of the Hungarian Academy of Sciences, H-1525 Budapest, Hungary}

\newcommand{\FIGWIDTH}{3.0in}

\begin{abstract}
  We argue that with the current experimental setup of the OPERA
  neutrino experiment no `bit' of information faster than light was or
  could be sent, and therefore no violation of Lorentz symmetry and/or
  causality was observed.

\end{abstract}

\maketitle

\section{INTRODUCTION}

In the recent OPERA neutrino experiment at Gran Sasso Laboratory,
superluminal propagation of neutrinos was reportedly observed
\cite{OPERA}. Since then lots of explanations of this surprising
result have emerged, e.g.~discussions on (the inconsistencies of)
tachyons \cite{drago}, on (the problematic relation with)
astrophysical data \cite{fargion}, on environmental explanation
\cite{tamburini}, on experimental details \cite{exp}, on general
relativity effects \cite{elburg}, etc.~(for a full list see the papers
citing \cite{OPERA}).

We assume in this note that the experiment is correct, so there is no
systematic shift in the measured data. Then one can ask whether it is
possible to send a bit of information faster than the speed of light
with this experimental setup, which would really mean violation of the
Lorentz symmetry or, even worse, of the well known causality
principle.

Apparent violation is appealing but not conclusive (see \cite{mecozzi}
for a related idea). In this respect it is interesting to note that
apparent superluminal phenomena exist, such as the famous EPR paradox
with superluminal `spooky' action at distance. However, up to now it
has proved impossible to send information faster than light.

\section{DISCUSSION}

\begin{figure}[ptb]
  \begin{center}
    \includegraphics[ width=\FIGWIDTH ]{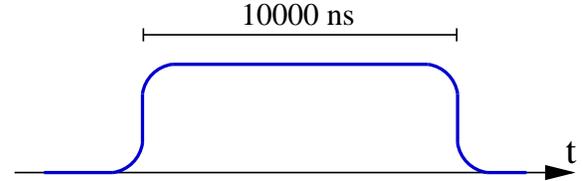}
  \end{center}
  \caption{Schematic form of the PDF as prepared in the sender located
    in CERN.}%
  \label{fig:prot_pdf}%
\end{figure}

During the experiment an approximately step-shaped proton distribution
function (PDF) was produced at CERN with time length of about
$10000$~ns, see Fig.~\ref{fig:prot_pdf} (and also Fig.~9 of
\cite{OPERA} for the precise form). Just after the measurement of the
PDF, the protons hit a target, resulting in the productions of the
mesons $\pi$ and $K$, which subsequently decay into muonic
neutrinos. The latter travel for about $d=730$ km in the inner earth
crust and reach the neutrino detectors of OPERA placed in Gran
Sasso. One expects that the distribution of muonic neutrinos has the
same (or very similar) shape of the original PDF. In this way it is
possible to measure the (mean) velocity of muonic neutrinos.

The finding is that the neutrino step function arrives about $60$ ns
earlier than expected. Technically, one has to shift by $1048.5$
ns the (scaled) PDF corresponding to the travel with the speed of
light $c.$ Then, one has to correct for $987.8$ ns, which is the
resulting time delay of various desynchronization effects. Finally,
one is left with the striking result that the neutrinos arrived
$60.7\pm13.5$ ns `before light', thus travelling at a velocity
$v_{\nu}>c$. For the following discussion it is crucial to stress that
the resulting time shift of $60$ ns, and therefore the conclusion that
$v_{\nu}>c$, is the result of a statistical analysis including the
leading as well as the trailing edges of the step functions. It is a
collective result based on the full length of the signal of about
$10000$ ns.

We turn now to the main question of this note: is it possible to send
a bit of information with the described OPERA experimental setup? One
may ask whether waiting only for the leading edge of the signal would
be enough (see the left plots of Fig.~12 of \cite{OPERA}). The leading
edge lasts about $1000$ ns, which is much larger than the $60$ ns time
difference. Restricting to these data only, it is very hard to see any
time shift. Moreover, the data oscillate and it would be impossible to
agree on the exact time of the incoming of the bit. More in general,
it seems impossible to send a bit of information faster than light
using only a very small portion of the PDF.

In the next section we elaborate in more detail, by considering a
simple modification of the PDF, on the possibility to send a bit of
information.

\section{A GEDANKENEXPERIMENT}

\begin{figure}[ptb]
  \caption{Coordinate-space profile of the neutrino packet needed to
    send one bit. The dashed lines correspond to sending either a zero
    or a one.}%
  \label{fig:bit_encode}
  \begin{center}
    \includegraphics[ width=\FIGWIDTH ]{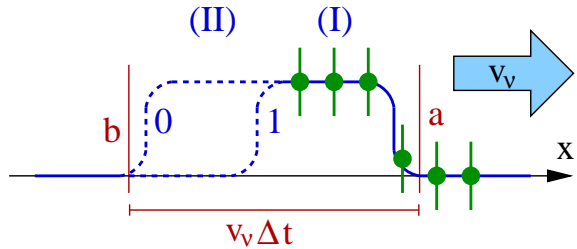}
  \end{center}
\end{figure}

Suppose we want to use the CNGS neutrino stream to transfer one bit,
such that the \textit{information} arrives earlier than it would have
had with standard light-based communication. One would then need to
shape the PDF (and the subsequent travelling neutrino packet) as
exemplified in Fig.~\ref{fig:bit_encode} (we assume that the proton
beam can be blocked at will). Over a background without neutrinos, a
bit-carrying signal is composed by two parts: part (I) triggers the
receiver, warning it that the bit will be encoded in part (II). This
two-part structure is necessary because the receiver does not know in
advance when exactly will the information be sent.

An important observation is that, while the ``button'' at CERN is
pressed corresponding to point $a$ in Fig.~\ref{fig:bit_encode}, the
bit will be completely received only at point $b$, that is, after the
whole sequence has been decoded, neglecting processing times and
similar practicalities. Thus the whole bit takes a somewhat large time
$\Delta t$ to be transmitted.

Moreover, the duration of parts (I) and (II) have to be previously
agreed upon by both sides. Denoting the time-resolution of the
receiver by $\Delta\tau=50$ ns, we can write $\Delta t=2N\Delta\tau$,
with $N$ the (smallest) number of data points necessary to identify
unambiguously the onset of a signal. Looking at the error bars and
slope of the curves in Fig.~12 of \cite{OPERA} (also sketched here in
Fig.~\ref{fig:bit_encode}), we give a conservative estimate $N\geq3$.

In order to achieve a truly faster-than-light transmission of a bit,
one needs even the \textit{tail} (point $b$) of the packet to arrive
faster than a light signal, shot at moment $a$, hence
\begin{equation}
  \Delta t=2N\Delta\tau<d\left(  \frac{1}{c}-\frac{1}{v_{\nu}}\right)
  \;\;,\label{req}%
\end{equation}
In the real-life case of the OPERA setup, the right-hand side is $60$
ns, which makes it impossible to fulfill the above requirement.

One may argue that a simple one-pulse signal (much like part (I) of
Fig.~\ref{fig:bit_encode} alone) would already imply violation of
Lorentz symmetry (or causality). This situation just amounts to
dropping the factor $2$ in Eq.~\ref{req}, but the conclusions would
remain unchanged. Moreover, it should also be noted that receiving a
neutrino in the detector implies a further time delay before the data
are received by the observer. This is contrary to receiving a photon,
where the use of scintillators allow for an (almost) instantaneous
processing: taking this into account amounts in a further increase in $N$.

Finally, it is amusing to imagine the existence of a straight tunnel
connecting CERN and Gran Sasso. This tunnel would be helpful to send,
starting at the same time $t=t_{a}$, a standard light-bit. In this way
it could be possible to check which \emph{signal} arrives
first. Unfortunately, as (almost) everyone knows, the construction of
such a tunnel is impossible for the moment (the price would surely
overshoot $45\cdot10^{6}$ euros).

\section{CONCLUSION}

In this work we have studied if it is possible to send a bit of
information with speed larger than $c$ with the experimental setup of
OPERA. Our result is negative: at present, this does not seem to be
feasible.

Our arguments are based on the particular present setup and therefore
one may object that, by improving the luminosity and the precision of
the experiment, or putting the neutrino detectors much further away,
the sending of a signal faster than $c$ would be indeed
possible. However, this represents an extrapolation which is based on
the assumption that nothing will change when doing that. Due to the
fact that Lorentz symmetry and causality are basic principles of our
understanding of nature, we believe that the rejection of (at least)
one of them should be motivated by the actual transmission of a bit of
information faster than $c$ and not by an extrapolation: Nature may be
subtle, protecting these basic principles with the help of some not
yet understood `censorship' mechanism.

\bigskip

\textbf{Acknowledgement}

The authors thank G.~Torrieri and G.~Pagliara for useful discussions.


\begin{thebibliography}{9} %

\bibitem {OPERA}T.~Adam \textit{et al.}~[OPERA Collaboration],
  [arXiv:1109.4897 [hep-ex]].

\bibitem {drago}A.~Drago, I.~Masina, G.~Pagliara, R.~Tripiccione,
  [arXiv:1109.5917 [hep-ph]]; A.~G.~Cohen, S.~L.~Glashow,
  [arXiv:1109.6562 [hep-ph]].

\bibitem {fargion}D.~Fargion,
  [arXiv:1109.5368 [astro-ph.HE]]; D.~Autiero, P.~Migliozzi, A.~Russo,
  [arXiv:1109.5378 [hep-ph]].

\bibitem {tamburini}F.~Tamburini, M.~Laveder,
  [arXiv:1109.5445 [hep-ph]].

\bibitem {exp}R.~Alicki,
  [arXiv:1109.5727 [physics.data-an]]; G.~Henri,
  [arXiv:1110.0239 [hep-ph]]; J.~Knobloch,
  [arXiv:1110.0595 [hep-ex]].

\bibitem {elburg}R.~A.~J.~van Elburg,
  [arXiv:1110.2685 [gen-ph]].

\bibitem {mecozzi}A.~Mecozzi, M.~Bellini,
  [arXiv:1110.1253 [hep-ph]].

\end{thebibliography}
\end{document}